\documentclass[prl,twocolumn,showkeys,showpacs,amsfonts,amssymb,amsmath]{revtex4}

\usepackage{graphicx}

\newcommand{\D}[2]{\frac{\partial #1}{\partial #2}}

\begin{document}

\title{Dynamics of Large-Scale Plastic Deformation\\
and the Necking Instability in Amorphous Solids}

\author{L. O. Eastgate}
\affiliation{Laboratory of Atomic and Solid State Physics, Cornell University,
Ithaca, New York 14853}
\altaffiliation{Temporary address: Department of Physics, University of
California, Santa Barbara, CA 93106.}
\author{J. S. Langer}
\author{L. Pechenik}
\affiliation{Department of Physics, University of California, Santa Barbara, CA
93106}
\date{August, 2002}

\begin{abstract}
We use the shear transformation zone (STZ) theory of dynamic plasticity to study
the necking instability in a two-dimensional strip of amorphous solid.  Our
Eulerian description of large-scale deformation allows us to follow the
instability far into the nonlinear regime.  We find a strong rate dependence;
the higher the applied strain rate, the further the strip extends before the
onset of instability. The material hardens outside the necking region, but the
description of plastic flow within the neck is distinctly different from that of
conventional time-independent theories of plasticity.
\end{abstract}

\pacs{62.20.Fe, 46.35.+z, 83.60.Df, 46.05.+b}

\keywords{plasticity, necking, numerics, STZ}


\maketitle

Conventional descriptions of plastic deformation in solids consist of
phenomenological rules of behavior, with qualitative distinctions between
time-independent and time-dependent properties, and sharply defined yield
criteria. Plasticity, however, is an intrinsically dynamic phenomenon.
Practical theories of plasticity should consist -- not of intricate sets of
rules -- but of equations of motion for material velocities, stress fields, and
other variables that might characterize internal states of solids.  Roughly
speaking, a theory of plasticity, especially for an amorphous solid, should
resemble the Navier-Stokes equation for a fluid, with the pressure replaced by a
stress tensor, and the viscous forces replaced by a constitutive law relating
the rate of plastic deformation to the stresses and internal state
variables. That constitutive law should contain phenomenological constants,
analogous to the bulk and shear viscosities, that are measurable and, in
principle, computable from molecular theories.  Yield criteria, work hardening,
hysteretic effects, and the like would emerge naturally in such a formulation.

The goal of the STZ (shear-transformation-zone) theory of plasticity
\cite{FALK:1998, LOBKOVSKY:1998, FALK:1999, LANGER:1999, FALK:2000, LANGER:2000,
LANGER:2001}, from its inception, has been to carry out the above program.  In
this paper we show how the STZ theory describes a special case of large-scale
yielding, specifically, the necking instability of a strip of material subject
to tensile loading. There is a large literature on the necking
problem. References that we have found particularly valuable include papers by
Hutchinson and Neale \cite{HUTCHINSON:1977}, McMeeking and Rice
\cite{MCMEEKING:1975}, and Tvergaard and Needleman \cite{TVERGAARD:1981}. Our
purpose here is to explore possibilities for using the STZ theory to investigate
a range of failure mechanisms in amorphous solids, possibly including fracture.
We are able to follow the necking instability far into the nonlinear regime
where the neck appears to be approaching plastic failure while the outer regions
of the strip become hardened and remain intact.  We find that necking in the STZ
theory is rate dependent; the instability occurs at smaller strains when the
strip is loaded slowly. One especially important element of our analysis is our
ability to interpret flow and hardening in terms of the internal STZ variables.

To make this problem as simple as possible, we consider here only strictly
two-dimensional, amorphous materials. By ``strictly,'' we mean that elastic and
plastic displacement rates are separately planar as in two-dimensional molecular
dynamics simulations. The two-dimensional STZ-equations presented in this paper
are based on earlier work by Falk, Langer, and Pechenik
\cite{FALK-THESIS,LANGER:2001,PECHENIK:2001}. We use Eulerian coordinates in
which, as in fluid dynamics, the variables $x_i$ denote the current physical
positions of material elements. Let the system lie in the $x_1=x$, $x_2 = y$
plane, and write the stress tensor in the form:
$\sigma_{ij}=-p\,\delta_{ij}+s_{ij},~~~p=-\frac{1}{2}\,\sigma_{kk}$, where $p$
is the pressure and $s_{ij}$ is the deviatoric stress -- a traceless, symmetric
tensor. In analogy to fluid dynamics, let $v_i(x,y,t)$ denote the material
velocity at the physical position $x,y$ and time $t$.  Then the acceleration
equation is: \cite{MALVERN} \begin{equation} \label{acceleration}
\rho\,\frac{dv_i}{ dt}=\D{\sigma_{ij}}{x_j}=-\D{p}{x_i}+\D{s_{ij}}{x_j}.
\end{equation} Here, $\rho$ is the density which, because we shall assume a very
small elastic compressibility and volume conserving plasticity, we shall take to
be a constant.  The symbol $d/dt$ denotes the material time derivative acting on
a scalar or a vector field:
\begin{equation}
\frac{d}{dt}\equiv \D{}{t}+v_k\,\D{}{x_k}.
\end{equation}

Our first main assumption is that the rate of deformation tensor can be written
as the sum of  linear elastic and plastic contributions:
\begin{align}
\label{Dtensor}
D^{\text{total}}_{ij}
&\equiv\frac{1}{2}\,
\left(\D{v_i}{x_j}+\D{v_j}{x_i}\right)\notag\\
&=\frac{\mathcal{D}}{\mathcal{D}t}\,
\left[\frac{s_{ij}}{2\mu} - \frac{p}{2K}\,\delta_{ij}\right]
+ D^{\text{plast}}_{ij},
\end{align}
where $\mu$ is the shear modulus, $K=\mu(1+\nu^*)/(1-\nu^*)$ is the
two-dimensional inverse compressibility (or bulk modulus), and $\nu^*$ is the
two-dimensional Poisson ratio. The symbol $\mathcal{D}/\mathcal{D}t$ denotes the
material time derivative acting on any tensor, say $A_{ij}$:
\begin{equation}
\frac{\mathcal{D}A_{ij}}{\mathcal{D}t}
\equiv \D{A_{ij}}{t}+v_k\,\D{A_{ij}}{x_k}+ A_{ik}\,
\omega_{kj} -\omega_{ik}\,A_{kj};
\end{equation}
and $\omega_{ij}$ is the spin:
\begin{equation}
\omega_{ij}=\frac{1}{2}\,\left(\D{v_i}{x_j}-\D{v_j}{x_i}\right).
\end{equation}

The plastic part of the rate-of-deformation $D^{\text{plast}}_{ij}$, like
$s_{ij}$, is a traceless symmetric tensor, thus the plastic deformations are
area-conserving.  For present purposes, we use a simple, quasi-linear form of
the STZ theory in which \begin{equation} \label{Dplastic} D^{\text{plast}}_{ij}=
\epsilon_0\,q_{ij}(s,\Delta); \quad\qquad q_{ij}(s,\Delta)= s_{ij}-\Delta_{ij},
\end{equation} and $\epsilon_0$ is a material-specific constant. The traceless,
symmetric tensor $\Delta_{ij}$ is the internal state variable mentioned earlier.
It is proportional to a director matrix that specifies the orientation of the
STZs; its magnitude is a measure of the degree of their alignment. The equation
of motion for $\Delta_{ij}$ is: \begin{equation} \label{Deltaeqn}
\frac{\mathcal{D}\Delta_{ij}}{\mathcal{D}t} =
q_{ij}-\frac{1}{2}\,|q_{km}\,s_{km}|\,\Delta_{ij}; \end{equation} In
Eq.~(\ref{Dplastic}), $\Delta$ plays -- very roughly -- the role of the ``back
stress'' or ``hardening'' parameter in conventional theories of
plasticity \cite{LUBLINER, KREMPL:1996, ABOUDI:1980},
a major difference being that $\Delta$ emerges directly from a rate equation
governing the population of STZs and is, in principle, a directly measurable
quantity \cite{FALK:1998,GAGNON:2001}. If the second term on the right-hand side
of Eq.~(\ref{Deltaeqn}) were missing, then $\Delta$ would be proportional to the
integrated plastic strain.  This second term, however, which is produced by the
creation and annihilation of STZs, is a crucial element of the STZ theory.  As
we shall show briefly below, this term produces the exchange of dynamic
stability between viscoelastic and viscoplastic states that replaces the
conventional assumptions of yield surfaces and other purely phenomenological
rules of behavior.
 
With one important exception, Eqs.~(\ref{Dplastic}) and (\ref{Deltaeqn})
constitute a tensorial version of the original STZ theory obtained by
linearizing the stress dependence of the rate factors and rescaling.  Because of
the linearization, these equations do not properly describe memory effects
present in the full theory that are important when the system is unloaded or
reloaded, but this will not affect our results until the system reaches the
necking instability. Only after the neck starts to flow plastically, causing the
hardened regions to unload, will we need the full non-linear theory to determine
if the observed behavior is pertinent.  We have chosen the rescaling so that all
stresses and moduli are expressed in units of the plastic yield stress.  We also
have assumed that the local density of STZs is always at its equilibrium value
so that we do not need to solve an extra equation of motion for that field
(denoted by the symbol $\Lambda$ in earlier papers).

The important exception alluded to above is the presence of the absolute-value
bars in Eq.~(\ref{Deltaeqn}). The expression inside the bars is proportional to
the rate at which plastic work is being done on the system, a quantity which
appears in the original theory as a non-negative factor in the STZ annihilation
and creation rates.  A negative value of this quantity would be unphysical.  In
earlier studies of spatially uniform systems, this quantity always remained
positive; however, we have observed negative values in the present
calculations. The absolute value prevents such unphysical behavior and is
consistent with the intent of the original theory. We emphasize, however, that
this term contains some of the principal assumptions of the STZ theory. There
are other possibilities for it (see, for example, \cite{LOBKOVSKY:1998}) and, as
yet, there is no first-principles derivation.

To understand the transition between viscoelastic and viscoplastic behaviors at
the yield stress, and the role played by the state variable $\Delta$, it is
easiest to look first at a uniform system under pure shear.  Let $s_{xx} =
-s_{yy} = s$, $s_{xy}=0$, $\Delta_{xx}=-\Delta_{yy}=\Delta$, $\Delta_{xy}=0$;
and consider a situation in which $s$ is held constant. Eqs.~(\ref{Dtensor}) and
(\ref{Deltaeqn}) become
\begin{align}
\dot\varepsilon &= \epsilon_0\,(s-\Delta)\\
\label{Deltaeqn2}
\dot\Delta &= (s-\Delta)(1-s\Delta),
\end{align}
where $\dot\varepsilon$ is the total strain rate.  At $s=1$, these equations
exhibit an exchange of stability between the non-flowing steady-state solution
with $\dot\varepsilon = 0$, $\Delta = s$ for $s<1$ and the flowing solution with
$\dot\varepsilon \ne 0$, $\Delta = 1/s$ for $s>1$.  As explained in earlier
publications, the steady-state system is ``jammed'' or ``hardened'' in the
direction of the applied stress for $s<1$ ;  whereas, for $s>1$, new STZs are
being created as fast as existing ones transform, and there is a nonzero plastic
strain rate.

Our goal now is to see how this exchange of stability occurs in a dynamic, spatially nonuniform situation.  Consider a rectangle with straight grips at $x=\pm L(t)$. The upper and lower surfaces, at $y = \pm Y(x,t)$, are free boundaries. We assume symmetry about both the $x$ and $y$ axes so that we need to consider only the first quadrant of the system. On the free upper boundary, the relation between the material velocities and the motion of the surface is 
\begin{equation}
\label{dYdt}
\left(\D{Y}{t}\right)_x = v_y(x,Y,t)-v_x(x,Y,t)\left(\D{Y}{x}\right)_t.
\end{equation}
We also must specify stress conditions on this surface:
\begin{equation}
\label{sigmann}
\sigma_{nn}=\gamma\,\mathcal{\kappa}
= \frac{\gamma\,Y^{\prime\prime}}{\left(1+Y^{\prime 2}\right)^{3/2}};~~~\sigma_{nt}=0.
\end{equation}
Here, $\gamma$ is the surface tension, $\kappa$ is the curvature, and the
subscripts $n$ and $t$ denote normal and tangential components respectively. The
grips at $x=\pm L(t)$ move outward at a predetermined strain rate, $\dot L/L =
\Omega$; thus $v_x(L,y,t)= L\,\Omega$ for $0 < y < Y(L,t)$. Note that we do not
constrain $v_y$ along this edge; we allow the grip to slide in the $y$
direction.


We wish to study how the shape of the upper surface, $Y(x,t)$, changes as the
grips on the sides are moved outward at various strain rates $\Omega$.  Rather
than trying to track this surface through the most general possible
deformations, we assume that $Y(x,t)$ remains single valued and simply make a
change of variables:
\begin{equation}
\xi = \frac{x}{L(t)},~~~~\zeta = \frac{y}{Y(x,t)}.
\end{equation}
We then transform Eqs.~(\ref{acceleration}), (\ref{Dtensor}), and
(\ref{Deltaeqn}), and the boundary conditions (\ref{dYdt}) and (\ref{sigmann})
into equations of motion for the velocity, the stress, the state variable
$\Delta$, and the moving boundary $Y$, all expressed as functions of $\xi$,
$\zeta$ and $t$.  We solve these equations in the fixed square $0<\xi<1$,
$0<\zeta<1$.

In all of the calculations described here, we have used $\rho=1$, $\mu=100$,
$K=300$, and $\gamma = 0.1$. Our initial conditions are $L(0)=4$ and $Y(x,0)=1 -
\delta(x)$, where $\delta(x)=0.01\exp(-8x^2)$ is a small deformation that breaks
translational symmetry.  We chose two values for $\epsilon_0$: 0.1 (hard) and
0.3 (soft), and two values for the strain rate $\Omega$: 0.01 (fast) and 0.001
(slow). The time taken by sound waves to cross the system is approximately
$L\,\sqrt{\rho/\mu}\approx 0.4$. This is smaller than the characteristic time
scale for plastic deformation, which we have scaled to unity, and is much
smaller than the actual time scales that we observe for our relatively small
pulling rates $\Omega$. Thus, our system is elastically quasi-stationary, and
the precise value of $\rho$ is not important.

We have solved these equations on a fixed, non-uniform $80\times 20$ grid in
$\xi,~\zeta$ space, using the implicit differential-algebraic solver DASPK
\cite{BROWN:1994}.  In order to suppress numerical instabilities, we have added
a small viscosity $\rho\eta\,\nabla^2\,v_i$ to the right-hand side of the
acceleration equation (\ref{acceleration}), and have set $\eta = 0.1$.

\begin{figure}
      \includegraphics[height=0.9\columnwidth,angle=270]{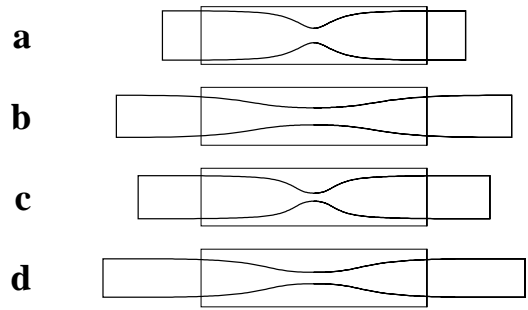}
      \caption{Initial and final shapes of the material in four numerical tensile tests:
      (a)~$\epsilon_0=0.1$, $\Omega=0.001$ (b)~$\epsilon_0=0.1$, $\Omega=0.01$
      (c)~$\epsilon_0=0.3$, $\Omega=0.001$ (d)~$\epsilon_0=0.3$, $\Omega=0.01$.}
      \label{fig:shapes}
\end{figure}

\begin{figure}
    \includegraphics[height=0.9\columnwidth,
    bb=81 0 500 681,angle=270]{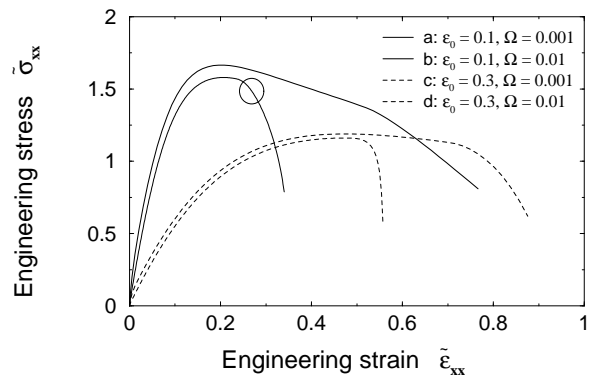}
    \caption{The engineering stress $\tilde{\sigma}_{xx}$ at the grip plotted
    against the engineering strain $\tilde{\varepsilon}_{xx}$ for the four cases
    shown in Fig.~\ref{fig:shapes}. The big circle marks the state whose
    internal properties are shown in Fig.~\ref{fig:s_delta}.}
    \label{fig:stress_strain}
\end{figure}

Fig.~\ref{fig:shapes} shows initial and final shapes of samples undergoing
tensile tests for four different combinations of the two parameters $\Omega$ and
$\epsilon_0$ as indicated. For clarity, we show the complete strip although only
the the behavior of the upper right quarter was computed. The final shapes were
arbitrarily chosen at the time when the engineering stresses at the grips were
roughly half of their peak values (see Fig.~\ref{fig:stress_strain}).
There is a necking instability in all four cases, but it occurs at greater
strain for the faster pulls. This rate dependence is also apparent in
Fig.~\ref{fig:stress_strain},
which shows the engineering stress at the center of the grip,
$\tilde{\sigma}_{xx}(L,0,t)=\sigma_{xx}(L,0,t)Y(L,t)/Y(L,0)$ as a function of
the engineering strain $\tilde{\varepsilon}_{xx}=[L(t)-L(0)]/L(0)$ for all four
cases (remember that $\sigma_{xx}=s-p$, and $s\equiv s_{xx}$).
Thus, although the ``softness'' parameter $\epsilon_0$ controls the overall
plastic response of the material, the onset of the necking instability is
controlled by the applied strain rate.

\begin{figure}
    \includegraphics[height=0.9\columnwidth,
    bb=81 0 492 660,angle=270]{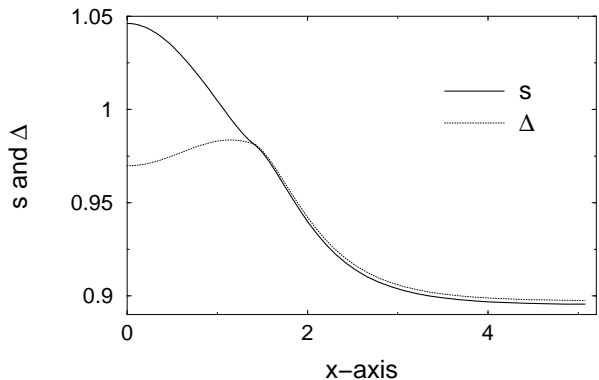}
    \caption{The deviatoric stress $s$, and $\Delta$, along the
    $x$-axis. This picture corresponds to case (a) indicated by the big circle
    in Fig.~\ref{fig:stress_strain} shortly after the sample is starting to
    neck.}
    \label{fig:s_delta}
\end{figure}

To see what is happening internally, we show in Fig.~\ref{fig:s_delta} graphs of
$s$ and $\Delta$($\equiv\Delta_{xx}$) along the centerline of the strip (the
$x$-axis) for case (a) in Fig.~\ref{fig:stress_strain} shortly after the sample
is starting to neck.
According to Eq.~(\ref{Dplastic}), the plastic flow rate is proportional to
$s-\Delta$.  Outside the necking region, $s\cong\Delta\le 1$; thus the system in
this region has hardened and deforms only elastically. Inside the necking
region, however, $s$ rises well above unity and $\Delta$ becomes small.  Here
the system has come close to steady-state flow on the $\Delta = 1/s$ branch of
stationary solutions of Eq.~(\ref{Deltaeqn2}).

This internal structure of the STZ picture of necking dynamics makes it clear
that the strain-rate dependence shown in Figs.~\ref{fig:shapes} and
\ref{fig:stress_strain} is caused by the competition between the rate of elastic
loading and the rate at which hardening occurs, the latter being governed by the
equation of motion for $\Delta$, Eq.~(\ref{Deltaeqn}).  When the loading is slow,
$\Delta$ grows along with the stress $s$, and there is little plastic flow
anywhere until the stress exceeds the yield stress in the necking region.  In
the opposite limit, when the loading is fast, $\Delta$ remains appreciably
smaller than $s$ for a longer time during which the material undergoes plastic
deformation everywhere.  It would be useful to test this prediction of the STZ
theory by measuring necking, say, in amorphous metals.  We presume that various
ingredients of the full STZ theory, such as stress-dependent rate factors and
other features that have been ignored here, would be needed to fit experimental
data quantitatively, and that we would learn much about the theory from such an
effort.

Note that the behavior shown in Fig.~\ref{fig:s_delta} is quite different from
that predicted by conventional, time-independent plasticity theory, in which
there would be a plastic zone with $s\cong 1$ inside the neck, that is, $s$
would remain at the yield stress.  We have found no evidence that this
conventional behavior occurs in the simulations presented here, even for the
smallest pulling speeds.  (For cavitation, the STZ theory predicts a
conventional plastic zone around a growing hole when the growth rate is very
slow \cite{LANGER:1999}.)  Once the instability sets in, the development of the
neck is governed by the elastic energy already stored in the strip.  We have
confirmed this feature of late-stage necking dynamics by performing numerical
experiments in which we stop the motion of the grips, that is, hold them fixed,
at various times after the neck has started to form but well before it has grown
appreciably.  We find that so long as the stored elastic energy is large enough,
stopping the remote loading in this way has almost no effect on the neck; it
continues to grow just as before, driven by the elastic unloading.

The behavior described in the last sentence -- necking driven by stored elastic
energy -- looks in many ways like fracture, although necking differs from
ordinary fracture in that the stress concentration that triggers the instability
is due to narrowing of the strip as a whole rather than to a localized defect on
just one surface. Nevertheless, the behaviors shown in Fig.~\ref{fig:s_delta}
(and other, later-stage results not shown here) suggest the onset of a
localized, propagating failure mechanism.  In order to study the connection
between necking and fracture in adequate detail we believe that we shall need to
use the full STZ theory and to improve our numerical resolution.

\begin{acknowledgments}
This research was supported primarily by U.S. Department of Energy Grant
No. DE-FG03-99ER45762. It was also supported in part by the MRSEC Program of the
NSF under Award No. DMR96-32716 and by a grant from the Keck Foundation for
Interdisciplinary Research in Seismology and Materials Science. Lance Eastgate
has been supported by the Research Council of Norway. We thank L. Petzold,
R. McMeeking, A. Needleman, and J. Rice for useful advice.
\end{acknowledgments}

\bibliography{allrefs}

\begin{thebibliography}{18}
\expandafter\ifx\csname natexlab\endcsname\relax\def\natexlab#1{#1}\fi
\expandafter\ifx\csname bibnamefont\endcsname\relax
  \def\bibnamefont#1{#1}\fi
\expandafter\ifx\csname bibfnamefont\endcsname\relax
  \def\bibfnamefont#1{#1}\fi
\expandafter\ifx\csname citenamefont\endcsname\relax
  \def\citenamefont#1{#1}\fi
\expandafter\ifx\csname url\endcsname\relax
  \def\url#1{\texttt{#1}}\fi
\expandafter\ifx\csname urlprefix\endcsname\relax\def\urlprefix{URL }\fi
\providecommand{\bibinfo}[2]{#2}
\providecommand{\eprint}[2][]{\url{#2}}

\bibitem[{\citenamefont{Falk and Langer}(1998)}]{FALK:1998}
\bibinfo{author}{\bibfnamefont{M.~L.} \bibnamefont{Falk}} \bibnamefont{and}
  \bibinfo{author}{\bibfnamefont{J.~S.} \bibnamefont{Langer}},
  \bibinfo{journal}{Phys. Rev. E} \textbf{\bibinfo{volume}{57}},
  \bibinfo{pages}{7192} (\bibinfo{year}{1998}).

\bibitem[{\citenamefont{Lobkovsky and Langer}(1998)}]{LOBKOVSKY:1998}
\bibinfo{author}{\bibfnamefont{A.~E.} \bibnamefont{Lobkovsky}}
  \bibnamefont{and} \bibinfo{author}{\bibfnamefont{J.~S.}
  \bibnamefont{Langer}}, \bibinfo{journal}{Phys. Rev. E}
  \textbf{\bibinfo{volume}{58}}, \bibinfo{pages}{1568} (\bibinfo{year}{1998}).

\bibitem[{\citenamefont{Falk}(1999)}]{FALK:1999}
\bibinfo{author}{\bibfnamefont{M.~L.} \bibnamefont{Falk}},
  \bibinfo{journal}{Phys. Rev. B} \textbf{\bibinfo{volume}{60}},
  \bibinfo{pages}{7062} (\bibinfo{year}{1999}).

\bibitem[{\citenamefont{Langer and Lobkovsky}(1999)}]{LANGER:1999}
\bibinfo{author}{\bibfnamefont{J.~S.} \bibnamefont{Langer}} \bibnamefont{and}
  \bibinfo{author}{\bibfnamefont{A.~E.} \bibnamefont{Lobkovsky}},
  \bibinfo{journal}{Phys. Rev. E} \textbf{\bibinfo{volume}{60}},
  \bibinfo{pages}{6978} (\bibinfo{year}{1999}).

\bibitem[{\citenamefont{Falk and Langer}(2000)}]{FALK:2000}
\bibinfo{author}{\bibfnamefont{M.~L.} \bibnamefont{Falk}} \bibnamefont{and}
  \bibinfo{author}{\bibfnamefont{J.~S.} \bibnamefont{Langer}},
  \bibinfo{journal}{MRS Bull.} \textbf{\bibinfo{volume}{25}},
  \bibinfo{pages}{40} (\bibinfo{year}{2000}).

\bibitem[{\citenamefont{Langer}(2000)}]{LANGER:2000}
\bibinfo{author}{\bibfnamefont{J.~S.} \bibnamefont{Langer}},
  \bibinfo{journal}{Phys. Rev. E} \textbf{\bibinfo{volume}{62}},
  \bibinfo{pages}{1351} (\bibinfo{year}{2000}).

\bibitem[{\citenamefont{Langer}(2001)}]{LANGER:2001}
\bibinfo{author}{\bibfnamefont{J.~S.} \bibnamefont{Langer}},
  \bibinfo{journal}{Phys. Rev. E} \textbf{\bibinfo{volume}{64}},
  \bibinfo{pages}{011504} (\bibinfo{year}{2001}).

\bibitem[{\citenamefont{Hutchinson and Neale}(1977)}]{HUTCHINSON:1977}
\bibinfo{author}{\bibfnamefont{J.~W.} \bibnamefont{Hutchinson}}
  \bibnamefont{and} \bibinfo{author}{\bibfnamefont{K.~W.} \bibnamefont{Neale}},
  \bibinfo{journal}{Acta Metall.} \textbf{\bibinfo{volume}{25}},
  \bibinfo{pages}{839} (\bibinfo{year}{1977}).

\bibitem[{\citenamefont{McMeeking and Rice}(1975)}]{MCMEEKING:1975}
\bibinfo{author}{\bibfnamefont{R.~M.} \bibnamefont{McMeeking}}
  \bibnamefont{and} \bibinfo{author}{\bibfnamefont{J.~R.} \bibnamefont{Rice}},
  \bibinfo{journal}{J. Solids Structures} \textbf{\bibinfo{volume}{11}},
  \bibinfo{pages}{601} (\bibinfo{year}{1975}).

\bibitem[{\citenamefont{Tvergaard et~al.}(1981)\citenamefont{Tvergaard,
  Needleman, and Lo}}]{TVERGAARD:1981}
\bibinfo{author}{\bibfnamefont{V.}~\bibnamefont{Tvergaard}},
  \bibinfo{author}{\bibfnamefont{A.}~\bibnamefont{Needleman}},
  \bibnamefont{and} \bibinfo{author}{\bibfnamefont{K.~K.} \bibnamefont{Lo}},
  \bibinfo{journal}{J. Mech. Phys. Solids} \textbf{\bibinfo{volume}{29}},
  \bibinfo{pages}{115} (\bibinfo{year}{1981}).

\bibitem[{\citenamefont{Falk}(1998)}]{FALK-THESIS}
\bibinfo{author}{\bibfnamefont{M.}~\bibnamefont{Falk}}, Ph.D. thesis,
  \bibinfo{school}{UCSB} (\bibinfo{year}{1998}).

\bibitem[{\citenamefont{Pechenik et~al.}(2001)\citenamefont{Pechenik,
  Rabinowitz, and Langer}}]{PECHENIK:2001}
\bibinfo{author}{\bibfnamefont{L.}~\bibnamefont{Pechenik}},
  \bibinfo{author}{\bibfnamefont{D.~J.} \bibnamefont{Rabinowitz}},
  \bibnamefont{and} \bibinfo{author}{\bibfnamefont{J.~S.}
  \bibnamefont{Langer}}, in \emph{\bibinfo{booktitle}{Bull. Am. Phys. Soc.}}
  (\bibinfo{organization}{APS March Meeting}, \bibinfo{year}{2001}),
  vol.~\bibinfo{volume}{46}, p. \bibinfo{pages}{1023}.

\bibitem[{\citenamefont{Malvern}(1969)}]{MALVERN}
\bibinfo{author}{\bibfnamefont{L.~E.} \bibnamefont{Malvern}},
  \emph{\bibinfo{title}{Introduction to the Mechanics of a Continuous Medium}}
  (\bibinfo{publisher}{Prentice-Hall, Inc.}, \bibinfo{address}{New Jersey},
  \bibinfo{year}{1969}).

\bibitem[{\citenamefont{Lubliner}(1990)}]{LUBLINER}
\bibinfo{author}{\bibfnamefont{J.}~\bibnamefont{Lubliner}},
  \emph{\bibinfo{title}{Plasticity Theory}} (\bibinfo{publisher}{Macmillan
  Publishing Company}, \bibinfo{address}{New York}, \bibinfo{year}{1990}).

\bibitem[{\citenamefont{Krausz and Krausz}(1996)}]{KREMPL:1996}
\bibinfo{editor}{\bibfnamefont{A.~S.} \bibnamefont{Krausz}} \bibnamefont{and}
  \bibinfo{editor}{\bibfnamefont{K.}~\bibnamefont{Krausz}}, eds.,
  \emph{\bibinfo{title}{Unified Constitutive Laws of Plastic Deformation}}
  (\bibinfo{publisher}{Academic Press}, \bibinfo{address}{San Diego},
  \bibinfo{year}{1996}), chap.~\bibinfo{chapter}{6}, p. \bibinfo{pages}{281}.

\bibitem[{\citenamefont{Aboudi and Bodner}(1980)}]{ABOUDI:1980}
\bibinfo{author}{\bibfnamefont{J.}~\bibnamefont{Aboudi}} \bibnamefont{and}
  \bibinfo{author}{\bibfnamefont{S.~R.} \bibnamefont{Bodner}},
  \bibinfo{journal}{Int. J. Eng. Sci.} \textbf{\bibinfo{volume}{18}},
  \bibinfo{pages}{801} (\bibinfo{year}{1980}).

\bibitem[{\citenamefont{Gagnon et~al.}(2001)\citenamefont{Gagnon, Patton, and
  Lacks}}]{GAGNON:2001}
\bibinfo{author}{\bibfnamefont{G.}~\bibnamefont{Gagnon}},
  \bibinfo{author}{\bibfnamefont{J.}~\bibnamefont{Patton}}, \bibnamefont{and}
  \bibinfo{author}{\bibfnamefont{D.~J.} \bibnamefont{Lacks}},
  \bibinfo{journal}{Phys. Rev. E} \textbf{\bibinfo{volume}{64}},
  \bibinfo{pages}{051508} (\bibinfo{year}{2001}).

\bibitem[{\citenamefont{Brown et~al.}(1994)\citenamefont{Brown, Hindmarsh, and
  Petzold}}]{BROWN:1994}
\bibinfo{author}{\bibfnamefont{P.~N.} \bibnamefont{Brown}},
  \bibinfo{author}{\bibfnamefont{A.~C.} \bibnamefont{Hindmarsh}},
  \bibnamefont{and} \bibinfo{author}{\bibfnamefont{L.~R.}
  \bibnamefont{Petzold}}, \bibinfo{journal}{SIAM J. Sci. Comp.}
  \textbf{\bibinfo{volume}{15}}, \bibinfo{pages}{1467} (\bibinfo{year}{1994}).

\end{thebibliography}

\end{document}